\documentstyle[aps,multicol]{revtex}
\renewcommand{\narrowtext}{\begin{multicols}{2} \global\columnwidth20.5pc}
\renewcommand{\widetext}{\end{multicols} \global\columnwidth42.5pc}
 
\input{epsf.tex}
\def\inseps#1#2{\def\epsfsize##1##2{#2##1} \centerline{\epsfbox{#1}}}

\begin{document}
\draft

\title{``Smoke Rings'' in Ferromagnets} 

\author{N. R. Cooper} 

\address{Theory of Condensed Matter Group, Cavendish Laboratory,
Madingley Road, Cambridge~CB3~0HE, United Kingdom, and, \\ School of
Physics and Astronomy, University of Birmingham, Edgbaston, Birmingham
B15 2TT, United Kingdom.}

\date{16 September, 1998}

\maketitle

\begin{abstract} 

It is shown that bulk ferromagnets support propagating non-linear
modes that are analogous to the vortex rings, or ``smoke rings'', of
fluid dynamics. These are circular loops of {\it magnetic} vorticity
which travel at constant velocity parallel to their axis of symmetry.
The topological structure of the continuum theory has important
consequences for the properties of these magnetic vortex rings.  One
finds that there exists a sequence of magnetic vortex rings that are
distinguished by a topological invariant (the Hopf invariant).  We
present analytical and numerical results for the energies, velocities
and structures of propagating magnetic vortex rings in ferromagnetic
materials.

\end{abstract}



\narrowtext

In many situations ferromagnetic materials may be viewed as continuous
media, in which the state of the system is represented by a vector
field indicating the local orientation of the magnetisation.  The
dynamics of the ferromagnet then follow from the time evolution of
this vector field, which obeys a non-linear differential equation
known as the Landau-Lifshitz equation.  Representing the local
orientation of the magnetisation by the vector field
$\bbox{n}(\bbox{r},t)$ ($\bbox{n}$ is a three-component unit vector),
the Landau-Lifshitz equation\cite{landaulifshitz} takes the form
\begin{equation}
\label{eq:ll}
\rho J \frac{\partial n_i}{\partial t} = - \epsilon_{ijk}
n_j\frac{\delta E}{\delta n_k} 
\end{equation}
in the absence of dissipation.  $\rho$ is the density of magnetic
moments, each of angular momentum $J$, and $E$ is the energy, which is
a functional of $\bbox{n}(\bbox{r},t)$ and its spatial derivatives
(summation convention is assumed throughout this paper).

The Landau-Lifshitz equation has been the subject of numerous
studies\cite{baryakhtar,odell,malozemovbook,kosevich,mikeska}, owing
to its physical importance as a general description of ferromagnetic
materials, and to the rich mathematical properties that result from
its combination of non-linearity and non-trivial topology.  Of
particular interest are the solitons and solitary
waves\cite{rajaraman} that it has been found to support. In one
spatial dimension, the Landau-Lifshitz equation is integrable for
certain energy functionals and the complete set of solitons is
known\cite{mikhailov,mikhailovshabat}. In higher dimensions, the
equation is believed to be non-integrable for even simple energy
functionals\cite{mikhailov}, and the understanding of non-linear
excitations\cite{kosevich} is incomplete.

Here, we construct a novel class of solitary waves of
three-dimensional Landau-Lifshitz ferromagnets.  Our approach relies
on the conservation of the linear momentum
\begin{equation}
\label{eq:mtm}
P_\alpha  \equiv 2\pi \rho J \epsilon_{\alpha\beta\gamma} \int
\! d^3\bbox{r}\; r_\beta \Omega_\gamma(\bbox{r},t) , 
\end{equation}
where $\Omega_\alpha\equiv
\frac{1}{8\pi}\epsilon_{\alpha\beta\gamma}\epsilon_{ijk}
n_i\nabla_\beta n_j\nabla_\gamma n_k$ is the ``magnetic vorticity''.
This definition of momentum\cite{paptom} resembles the definition of
the hydrodynamic impulse of an incompressible fluid\cite{saffman}, if
fluid vorticity is identified with magnetic vorticity.  Our work
emphasises the connection to fluid dynamics by showing that there
exist solitary waves in ferromagnets that are analogous to the vortex
rings of fluid dynamics\cite{saffman,shariff}: these are circular
loops of (magnetic) vorticity that propagate at a constant velocity
parallel to their axis of symmetry.  (The magnetic analogues of
vortex/anti-vortex pairs have recently been determined using a similar
approach\cite{cooper2dsolitarywaves}.)

Furthermore, just as there exist
generalisations\cite{moffatt88,turkington} of the vortex rings in
fluid dynamics to vortex ring structures in which the lines of
vorticity are linked (as measured by a non-zero
helicity\cite{moffatt69}), there exist similar generalisations of the
magnetic vortex rings in ferromagnets to topologically non-trivial
structures involving the linking of vortex
lines\cite{dzylsoliton,papanicolaou_rings} (in this context the
measure of linking is known as the Hopf invariant\cite{botttu}).  As a
consequence, ferromagnets support a {\it sequence of
topologically-distinct magnetic vortex rings}.  One can understand how
this sequence arises by noting that a magnetic vortex line carries an
internal orientation that can be twisted through an integer multiple
of $2\pi$ as the vortex line traces out a closed
loop\cite{wilczekzee}.  By varying the number of rotations of this
internal angle, one obtains a sequence of magnetic vortex ring
configurations that are topologically distinct (they cannot be
interconverted by non-singular deformations), as they relate to
different values of the Hopf invariant\cite{botttu}.  (See
Ref.\onlinecite{moffatt90} for a similar construction for fluids,
where topology allows the inserted twists to be non-integer multiples
of $2\pi$; magnetic vortex rings are more akin to the coreless vortex
rings of superfluid $^3$He-$A$, Ref.\onlinecite{ho78}.)  The Hopf
invariant, ${\cal H}$, is the integer invariant that characterises the
mappings $S^3\rightarrow S^2$ (the vector field $\bbox{n}(\bbox{r})$
describes such a mapping when fixed boundary conditions are imposed,
e.g. $\bbox{n}(|\bbox{r}|\rightarrow\infty)=+\bbox{\hat{z}}$); it can
be interpreted in terms of the linking number of two vortex lines on
which $\bbox{n}$ takes different values\cite{botttu}.  This
topological invariant has been of interest in recent studies of
non-linear field theories, where it has been used to stabilise static
solitons\cite{faddeev,glad,battye3d}.  Here we show that it classifies
a sequence of {\it dynamical} solitary waves of ferromagnets -- the
magnetic vortex rings.

We shall construct magnetic vortex ring solitary waves for
ferromagnetic materials described by the energy functional
\begin{equation}
\label{eq:fun}
E \equiv \frac{1}{2} \rho_s \int
\! d^3\bbox r \; (\nabla_\alpha n_i)^2  + \frac{1}{2}A\int \! d^3\bbox{r} \; (1-n_z^2),
\end{equation}
which represents isotropic exchange interactions and uniaxial
anisotropy (we consider only $A\geq 0$, and choose the groundstate to
be the uniform state with $\bbox{n}=+\bbox{\hat{z}}$).  It is
straightforward to verify that, for this functional, the momentum
(\ref{eq:mtm}) is conserved by the dynamics (\ref{eq:ll}), as is the
number of spin-reversals
\begin{equation}
N \equiv  \rho J/\hbar \int \! d^3\bbox{r}\; (1-n_z) .
\label{eq:num}
\end{equation}
(Within a full quantum description, the number of spin-reversals would
be an integer; within the semiclassical description afforded by the
Landau-Lifshitz equation, which is accurate for $N\gg 1$, $N$ a
continuous variable.)  Our approach is to find configurations,
$\bbox{n}^*(\bbox{r})$, that extremise the energy (\ref{eq:fun}) at
given values of the momentum, $\bbox{P}$, and number of
spin-reversals, $N$, within each topological subspace, ${\cal
H}$. This procedure\cite{tjonwright,cooper2dsolitarywaves} defines an
extremal energy $E^*_{{\cal H}}(\bbox{P},N)$.  The variational
equations can be used to show that there exist time-dependent
solutions of Eqn.(\ref{eq:ll}) of the form
\begin{eqnarray*}
n_x(\bbox{r},t)+in_y(\bbox{r},t) & = & [n_x^*(\bbox{r}-\bbox{v} t)
+in_y^*(\bbox{r}-\bbox{v} t)]e^{i\omega t}\\ n_z(\bbox{r},t) & = &
n_z^*(\bbox{r}-\bbox{v} t)
\end{eqnarray*}
where
\begin{equation}
\label{eq:vel}
v_\alpha = \left.\frac{\partial E_{{\cal H}}^*}{\partial
P_\alpha}\right|_N \; ;\;\;\;\; \omega =
\left.-\frac{1}{\hbar}\frac{\partial E_{{\cal H}}^*}{\partial
N}\right|_{P_\alpha}.
\end{equation}
These solutions describe travelling waves which move in space at
constant velocity, $v_\alpha$, while the magnetisation precesses
around the $z$-axis at angular frequency $\omega$. We shall find
configurations $n_i^*(\bbox{r})$, resembling magnetic vortex rings,
which have spatially-localised energy density and therefore describe
propagating solitary waves\cite{rajaraman}.

It has been suggested previously that solitary waves resembling
magnetic vortex rings might exist in ferromagnets, being stabilised by
a non-zero Hopf invariant ${\cal H}$ combined with the conservation of
either the number of spin-reversals\cite{dzylsoliton} $N$, or the
momentum\cite{papanicolaou_rings} $\bbox{P}$. In the present work we
make use of the conservation of {\it both} $N$ and $\bbox{P}$,
allowing solitary waves with more general motions to be constructed.
This proves to be essential for the magnetic vortex rings that we
discuss here.  Furthermore, since we do not invoke topological
stability, we can obtain solitary waves for all values of the Hopf
invariant, including ${\cal H}=0$.

We now turn to the determination of the properties of magnetic vortex
ring solitary waves within the differing topological subspaces ${\cal
H}$.  First, we make some general remarks. Due to the invariance of
the energy (\ref{eq:fun}) under spatial rotations, the extremal energy
$E_{{\cal H}}^*(\bbox{P},N)$ is independent of the direction of
momentum; for later convenience, we choose
$\bbox{P}=P\hat{\bbox{y}}$. By considering the dependences of
Eqns.(\ref{eq:mtm},\ref{eq:fun},\ref{eq:num}) under scale
transformations, one finds that the extremal energy has the form
$E_{{\cal H}}^*(P,N)=N^{1/3}{\cal E}_{\cal H}(p,a)$, where $p\equiv
P/N^{2/3}$ and $a\equiv A N^{2/3}$ (here, and subsequently, we choose
units for which $\rho_s=\rho J=\hbar=1$).

As a first step, consider linearising the equations of motion about
the groundstate $\bbox{n}(\bbox{r})=\hat{\bbox{z}}$.  The
configurations that extremise the energy at fixed $N$ and $\bbox{P}$
are easily found: they are spatially-extended, and describe $N$
non-interacting spin-waves each of momentum $\bbox{P}/N$. The results
of this analysis determine the spin-wave dispersion ${\cal
E}^{sw}_{{\cal H}=0}(p,a) = p^2 + a$ (the Hopf invariant is zero for
all small-amplitude disturbances).

Insight into the results of the full, non-linear theory may be
achieved by considering large-radius magnetic vortex ring
configurations. Consider a magnetic vortex carrying a total ``flux''
$Q$, that is closed to form a circular loop with radius much larger
than the size of the vortex core. (The vortex core size, like the loop
radius, varies with $N$ and $P$. The condition that the core is small
compared to the radius is $P/QN^{2/3} = p/Q \gg 1$.)  The flux $Q$ is
defined to be the integral of the vorticity across a surface pierced
by the vortex; since the magnetisation tends to a constant,
$\bbox{n}=\bbox{\hat{z}}$, far from the vortex core, $Q$ is an integer
(this is the topological invariant that classifies the mappings
$S^2\rightarrow S^2$).  Neglecting the size of the vortex core in
comparison to the radius of the loop, one can determine the minimum
energy of these magnetic vortex rings using results from previous
studies of topological solitons in two-dimensional
ferromagnets\cite{kosevich}.  One finds
\begin{eqnarray}
\label{eq:asym1}
{\cal E}^{asym}_{\cal H}(p,a) & = & 4\pi\sqrt{|Q|p} +
a/|Q|,\;\;\;\;\; a^2\ll p/Q, \\ {\cal E}^{asym}_{\cal H}(p,a) & = &
2\sqrt{2\pi a}(p/|Q|)^{1/4}   ,\;\;\;\;\; a^2\gg p/Q ,
\label{eq:asym2}
\end{eqnarray}
which are valid for $p/Q\gg 1$, when the vortex core is small compared
to the radius of the vortex ring ($Q\neq 0$ has been assumed).  The
relative sizes of $a^2$ and $p/Q$ determine whether the vortex core is
small [Eqn.~(\ref{eq:asym1})] or large [Eqn.~(\ref{eq:asym2})]
compared to a lengthscale set by the anisotropy (the width of a domain
wall).

In order to determine the properties of the magnetic vortex rings when
the size of the vortex core is comparable to the radius of the ring
($p/Q\sim 1$) we have employed numerical analysis (we have not been able
to solve the variational equations analytically).  We simplify the
problem by minimising the energy within a class of axisymmetric
configurations\cite{glad} that is consistent with the general
variational equations:
\begin{eqnarray*} 
\left[n_x(\bbox{r}) + in_y(\bbox{r})\right] & = & 
\left[n_x(\rho,y)+in_y(\rho,y)\right]e^{iM\phi}, \;\;\;\;\;\;\;\; y\geq 0,  \\
\left[n_x(\bbox{r}) + in_y(\bbox{r})\right] & = & 
\left[n_x(\rho,-y)-in_y(\rho,-y)\right]e^{-iM\phi},  \; y\leq 0,  \\
n_z(\bbox{r}) & = & n_z(\rho,|y|), 
\end{eqnarray*}
where $(\rho,y,\phi)$ are cylindrical polar co-ordinates about the
$y$-axis.  We thereby reduce configuration space to the functions
$\bbox{n}(\rho,y)$ on the quarter plane $(\rho\geq 0, y\geq 0)$ and
the integer parameter $M$. The Hopf invariant of these configurations
is\cite{glad} ${\cal H}= M Q$, where $Q$ is the total flux through the
half-plane $(\rho\geq 0)$.  For $M\neq 0$, finite energy
configurations have $\bbox{n}\rightarrow \bbox{\hat{z}}$ at
$\rho\rightarrow 0$ as well as at $\rho^2+y^2\rightarrow\infty$, so
the total flux $Q$, and hence the Hopf invariant, take only integer
values.

We have studied the isotropic ferromagnet, $a=0$, by a discretisation
of the region $(0\leq\rho\leq L, 0\leq y\leq L/2)$ on square lattices
of up to $600\times 300$ sites. Fixed boundary conditions,
$\bbox{n}=\bbox{\hat{z}}$, were imposed on $\rho=L$ and $y=L/2$, and
$n_y=0$ was imposed on $y=0$ for consistency with the above ansatz
(the requirement that $\bbox{n}=\bbox{\hat{z}}$ on $\rho=0$ when
$M\neq 0$ emerges naturally from the energy minimisation).  Energy
minimisation was achieved by a conjugate gradient method with
constraints on $N$ and $P$ imposed by an augmented Lagrangian
technique\cite{practicaloptimization}.

\begin{figure}
\inseps{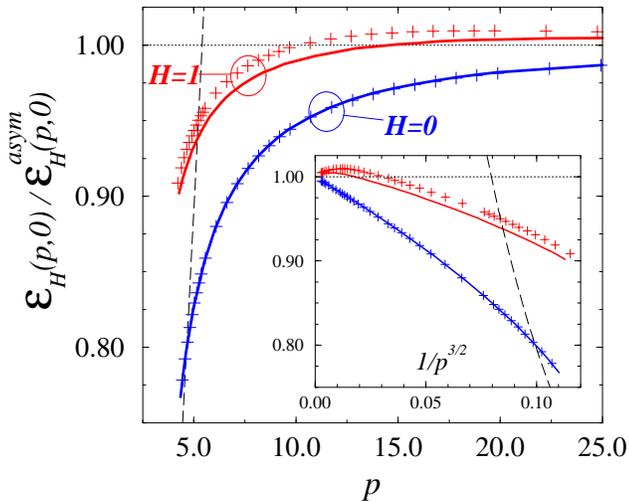}{0.47}
\caption{Energies of the magnetic vortex ring solitary waves with
${\cal H}=0$ and ${\cal H}=1$ for $a=0$ (shown as the ratio to the
asymptotic energy (\protect\ref{eq:asym1}) with $Q=1$).  To indicate
the extent of finite-size effects, two system sizes are shown: the
largest we have studied $L=11.2 N^{1/3}/p^{1/4}$ (solid lines), and
$L=8.2 N^{1/3}/p^{1/4}$ (crosses).  The spin-wave dispersion is shown
as a dashed line. Inset: the same quantities as functions of
$1/p^{3/2}$.}
\label{fig:disp}
\end{figure}

The energies of the magnetic vortex ring solitary waves with ${\cal
H}=0$ ($M=0$) and ${\cal H}=1$ ($M=Q=1$) are shown in
Fig.~\ref{fig:disp} as functions of the scaled momentum $p$.  At large
momenta, both branches approach the asymptotic expression
(\ref{eq:asym1}) for a $Q=1$ vortex. Assuming that the leading
corrections shown in the inset are linear in $1/p^{3/2}$, one finds
from Eqns.~(\ref{eq:vel},\ref{eq:asym1}) that a vortex loop of large
radius, $p/Q\gg 1$, moves in such a way that its precession frequency
$\omega$ is comparable to the frequency with which it translates a
distance of order its own radius.  At small momenta, the solitary
waves become higher in energy than non-interacting spin-waves for $p<
4.66$ (${\cal H}=0$) and $p< 5.1$ (${\cal H}=1$). Both branches are
found to persist below these values for a small range of $p$ as {\it
local} energy minima.

\begin{figure}
\inseps{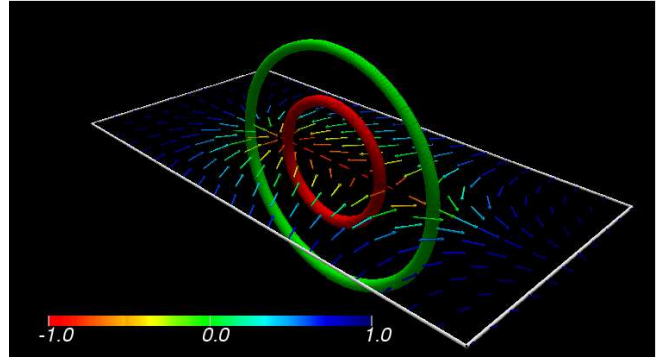}{0.45}
\vskip0.2cm
\caption{The configuration of the magnetic vortex ring with ${\cal
H}=0$, $p=5.0$, and $a=0$.  The vectors represent the projection of
$\bbox{n}$ onto the $xy$-plane; their colours indicate the value of
$n_z$ (refer to the colour-legend).  The tubular curves illustrate the
lines along which $\bbox{n}=-\bbox{\hat{z}}$ (red) and
$\bbox{n}=+\bbox{\hat{x}}$ (green). (For clarity, only the central
section of the full lattice is illustrated, with a fraction of the
lattice points reproduced.) }
\label{fig:h0}
\end{figure}

\begin{figure}
\vskip-0.2cm
\inseps{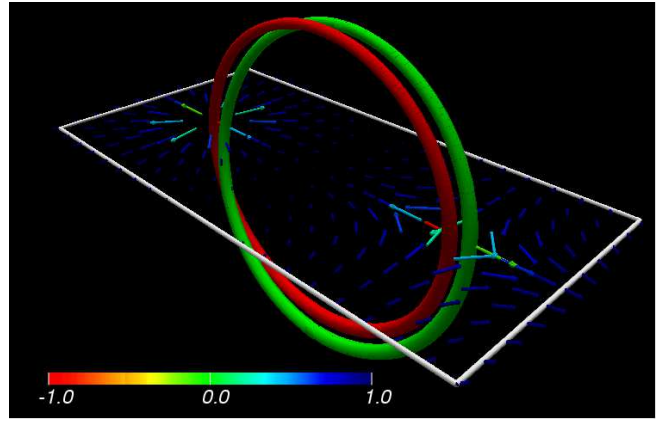}{0.45}
\caption{The same as Fig.\protect\ref{fig:h0}, for ${\cal H}=1$,
$p=5.3$, and $a=0$.}
\label{fig:h1}
\end{figure}

Typical configurations of the solitary waves are illustrated in
Figures~\ref{fig:h0} and~\ref{fig:h1}, which show the local
magnetisation within the $xy$-plane and three-dimensional
representations of the curves on which $\bbox{n}=-\bbox{\hat{z}}$ and
$\bbox{n}=+\bbox{\hat{x}}$.  These curves allow a direct visualisation
of the topological structure of the configurations, as measured by the
Hopf invariant\cite{botttu}. In Fig.~\ref{fig:h0}, the curves are
simply circles centred on the $y$-axis: they are unlinked,
illustrating that the Hopf invariant is zero, ${\cal H}=0$. In
Fig.~\ref{fig:h1}, the two curves are linked once, illustrating a
non-trivial topological configuration with unit Hopf invariant, ${\cal
H}=1$.


For larger values of the Hopf invariant (${\cal H}\geq 2$) we find
that corrections arising from the finite lattice size in our
calculations become more significant.  (Finite-size effects are
already apparent in Fig.~\ref{fig:disp} for ${\cal H}=1$.)  These
effects prevent a convincing demonstration of the existence of
(non-singular) magnetic vortex rings with ${\cal H}\geq 2$ that
describe stable energy minima. The construction of stable magnetic
vortex rings with ${\cal H}\geq 2$ may require the use of
non-axisymmetric configurations.  The results of the present study
demonstrate that magnetic vortex rings with ${\cal H}=0$ and ${\cal
H}=1$ do describe stable energy minima within the class of
axisymmetric configurations assumed.  Our results provide the
structures, energies, and therefore the velocities and precession
frequencies [see Eqns. (\ref{eq:vel})] of these propagating non-linear
modes.  Further numerical studies\cite{bigpaper} indicate that similar
magnetic vortex ring solitary waves exist for non-zero anisotropy,
$a\neq 0$. (In fact, the magnetic vortex rings are apparently more
favourable: e.g. the branch of solitary waves with ${\cal H}=0$
persists down to $p=0$ for $a\gtrsim 23.5$, consistent with the
existence of purely precessional solitary waves in a uniaxial
ferromagnet\cite{kosevich}.)  Any additional sources of magnetic
anisotropy in experimental systems, or the inclusion of magnetic
dipole interactions\cite{demagnet}, will lead to corrections that are
small when the magnetic vortex rings are sufficiently small (compared
to a characteristic lengthscale set by the strength of these
additional couplings).

One may wonder why magnetic vortex rings have not as yet been observed
experimentally, whilst vortex rings in fluids are a matter of everyday
experience.  The answer lies in the difficulty of creating non-linear
excitations in solid-state materials.  For example, magnetic vortex
rings involve a large number of spin-reversals, so they are not
accessed in standard inelastic neutron scattering experiments which
probe single spin-flip excitations. The creation of magnetic vortex
rings in ferromagnetic materials will require the use of other
experimental techniques.  One way in which magnetic vortex rings could
be excited experimentally, the details of which we are currently
investigating, is to exploit an instability that we have discovered to
the creation of magnetic vortex rings in itinerant ferromagnets of
mesoscopic size under conditions of high current density.  This
instability signals a transition to a form of magnetic turbulence in
mesoscopic ferromagnets, driven by the exchange coupling between the
magnetisation and the spins of the itinerant electrons, and may be
relevant to the unexplained dissipative phenomena observed in
Ref.\onlinecite{tsoi}.

\vskip0.3cm

\noindent
The author is grateful to Mike Gunn and Richard Battye for helpful
discussions, and to Pembroke College, Cambridge and the Royal Society
for financial support.


\begin{thebibliography}{10}
\bibitem{demagnet} Magnetic dipole interactions cause a
non-conservation\cite{paptom} of the number of spin-reversals, $N$,
which will limit the timescale over which modes involving precession
behave as solitary waves.  Whether there exist propagating magnetic
vortex rings that are stable even in the presence of dipolar
interactions\cite{papanicolaou_rings} remains an open question.
\end{thebibliography}

\widetext

\end{document}